# The Crucial Formula for Determination of the Occurrence of the Non-Chaotic States in the rf-biased Nonlinear Oscillators

Tsung Hsun Yang

*Institute of Electro-Optical Engineering, National Chiao-Tung University,*

*1001 Ta-Hsueh Road, Hsinchu, Taiwan 300, Republic of China*

Ching Sheu Wang

*Telecommunication Laboratories, Ministry of Transportation and*

*Communications, Taoyuan, Taiwan 300, Republic of China*

Jeun Chyuan Huang

*Communication Technology Division, Industrial Technology Research Institute,*

*Chung Chen Road, Hsinchu, Taiwan 300, Republic of China*

Yih Shun Gou

*Institute of Electrophysics, National Chiao-Tung University, 1001 Ta-Hsueh*

*Road, Hsinchu, Taiwan 300, Republic of China*

The crucial formulas to determine the non-chaotic states in the rf-biased nonlinear oscillators are derived from the numerical experiments. The nature of these formulas, which depends on symmetrical properties of the potential well, in terms of the driven-frequency as a function of the damping constant k is investigated. All these ones provide crucial guide posts to check which kinds of solutions (simple or complicated) can be tailored in the dissipative rf-biased nonlinear oscillators, respectively.

## 1. INTRODUCTION

In the nonlinear oscillating systems with an external driven force, the equation of motion can be expressed as a general form:

$$\ddot{x} + k\dot{x} + f(x) = F\sin\omega t \tag{1}$$

where $k$ denotes the damping constant, F and $\omega$ represent the amplitude and the frequency of the external periodic driving force respectively, and f(x) is the nonlinear term. If f(x) can be expressed as $x + x^2 + \cdots$, the Eq.(1) becomes as the dynamic motion of Duffing oscillators. Whereas if $f(x) = sinx$, the equation is called the resistant-shunted Josephson junction model. Following previous work[1-8], the complicated solutions to the Eq.(1), including period-$2^k$, chaotic, period-$3 \times 2^k$ and period-$m \times 2^k$ ones ($k$ is a positive integer and m a prime), have been formed in the parameter region where $\omega$ and $k$ are rather small. Moreover, the non-chaotic solutions to the Eq.(1), including symmetric and asymmetric period-1 ones, dominate in a manner of large $k$ and large $\omega$. In this case, no matter what $F$ is, there is only symmetry-breaking bifurcation observed. In fact, an occurrence of non-chaotic states was first indicated by McDonald and Plischke[9] in the study of the resistant-shunted Josephson junction (RSJ model[10,11]) with the following criterion:

$$1 << k^2 \tag{2}$$

i.e., the damping constant is much larger than a unit. And then subsequently Kautz[12,13] pointed out that chaos does not occur for

$$\omega >> k \tag{3}$$

According to their suggestions, there are only simple or stable states survivable in some confined region of the parameter space of the higher damping and/or the faster driving frequency cases. Although these criteria can enlighten the parameter regions of the complicated motion qualitatively, a quantitative resolution of this problem need explore yet. In other words, the results described above immediately raises an intriguing problem : can one exactly predict where or how a nonlinear oscillating system will display non-chaotic states in terms of chosen parameters $\omega$ and $k$ quantitatively.

To this end, herein, we will attempt to deduce certain crucial formulas in terms of the parameters, $\omega$ and $k$, to quantitatively distinguish the existence of chaos and the occurrence of non-chaotic states in Eq.(1) not only in the RSJ model but also in Duffing oscillators. In what follows the method of derivation of the formulas with the aid of the phase diagrams, bifurcation diagrams, state diagram and so on will be described. As compared with the RSJ model and the generalized Duffing oscillators, the physical implications of the symmetrical properties of potential wells related to the character of the formulas will be examined. Concluding remark will be reported as well.

## 2. The Numerical Simulation

As to derive the crucial formulas to distinguish the existence of chaos from the occurrence of non-chaotic states in the more general cases of the dissipative nonlinear oscillator, two typical nonlinear oscillating systems, RSJ model and Duffing oscillators, are considered at present. In the numerical simulation, the fourth order Runge-Kutta algorithm[14] is employed to do integration. Based on the integration, the Newton-Raphson method[14] is utilized to find the stable and unstable fixed points on the Poincaré sections. Two scanning procedures, by varying driving frequency at a fixed driving amplitude ($\omega$ scanning) and by varying driving amplitude at a fixed driving frequency ($F$ scanning), are employed with damping constant $k$ as a fixed parameter to get as much information as possible. Consequently, various characteristics of all the thresholds of possible transitions in terms of the controlled parameters can be obtained and depicted in a state diagram.

In each state diagram of the respective systems, for example as shown in Fig.1, there may exist a threshold frequency, $\omega_T$, at a fixed damping constant, $k$. We found that if the driving frequency is larger than the frequency, $\omega_T$, in this damping case, no matter how large the driving force amplitude is the chaotic solutions cannot exist. Then, as varying the damping constant $k$ to the other fixed value we can obtain another $\omega_T$ with the same process. Finally, to repeat the process again and again, the relations between $\omega_T$ and $k$ can be established in this way after all. These relations allow us to derive a border line and then to formulate a closed mathematical form in the $\omega - k$ space, by which the complicated solution region from the simple stable period-1 solution one can be differentiated. The detail contents of such $\omega - k$ relations of the formulas for the RSJ model and the Duffing oscillators will be discussed below.

## A. RSJ Model

The rf-biased RSJ model is given as:

$$\ddot{x} + k\dot{x} + \sin x = F \sin \omega t \tag{4}$$

where $k$ is the damping constant, $F$ is the amplitude of rf driving force and $\omega$ is the frequency of rf driving force. Eq.(4) describes the equation of motion of the phase difference of the junction. The equation can also be visualized as the dynamic motion of the swept angle of pendulum[9,10,15], charge density wave[9], a parametric amplifiers[16,17], et. al.. By means of the numerical method described above, the features of the state diagram related to various solutions of the states[9−21] in Eq.(4) will be presented as follows. For the sake of illustration, only seven branches (shaded regions) are denoted in the $F - \omega$ states diagram as $k = 1.0$ as shown in Fig.1. There are two kinds of solution regions. One contains the simple solution including the period-1 oscillatory (symmetric and symmetry-broken) and the phase-locked traveling ones. And the other contains the complicated solution including period-$2^k$, chaotic, period-$3 \times 2^k$, period-$m \times 2^k$ ($k$ is a positive integer and $m$ a prime) and the phase-unlocked traveling ones. The shaded region of the branches indicate the existence of the complicated solutions. Outside these regions, especially in the region where the frequency is above $\omega_T$, there does exist the simple solutions without observing the chaotic ones. In fact, the scenario of a sequence of bifurcation in each branch along the $F$-axis at a fixed frequency $\omega$, is mostly the same, as shown in Fig.2. They mainly contain period-doubling cascades, chaos, intermittency, crisis, and reverse period-doubling cascades. The period-3 windows and sometimes the period-5 windows very often burst out within the chaotic events. Since the similar scenarios of bifurcation appear itself again and again in each branch, it implies the self-similarity of the solutions embedded in the RSJ model clearly. The greater details of such self-similar properties will be seen subsequently.

As also shown in Fig.1, the territory of the each branch is bounded by the terrain of the symmetry-breaking (SB) solution, respectively. There are the regions of the symmetric solution existed between the regions of the SB solutions as well. Moreover, both the frequency of symmetric and SB solutions are period-1. The shaded regions of the branches will shrink toward smaller and smaller, closer and closer and then finally merge together when $\omega$ is decreasing and $F$ is increasing further and further. We have find that in the left-upper corner of the Fig.1, actually, there are still much more branches existed.

In addition to the evidences of the above observations, the self-similar feature of the bifurcation in each branch in the RSJ model can be further illuminated as follows. Firstly, as shown in Fig.1, the terrains of each simple solution separated by the branches of the complicated regions are marked on a series of consecutive integers respectively from down to up. We note that the integers just reveal the corresponding number of the local wells of the infinite sinusoidal potential within which the oscillating motion of the simple solution executes in the RSJ model. For example, in the case of $\omega = 0.313, F = 1.20$ and $k = 1.0$, the response is oscillating in the $n = 2$ regime as shown in Fig.1, and corresponds to the motion swinging over two local wells of the sinusoidal potential within the interval of one period of the rf-biasing as shown in Fig.3b. Similarly, the phase portraits, as shown in Fig.3, illustrate these behaviors apparently from $n = 1$ to $n = 7$, respectively, when $F$ is increasing from 1.80 to 7.0 for $\omega = 0.313$. The situation is further demonstrated from Fig.3g for $n = 51$ as well. In the complicated region, the common effect of the response swinging extensively from "$n$-th" local well to "$n+1$-th" one results from two stable solutions merging with an unstable one which located on the top of the sinusoidal well. In other words, the respective asymmetrical solutions in the adjacent wells are merging together into one solution. It leads to the swinging extensively one more well after the collision between the asymmetrical solutions and the unstable one.

It is also worthy noting that all the branches in the state diagram at a fixed damping constant $k$ are located within the region where the driving frequency is less than the critical threshold frequency $\omega_T$. If the driving frequency is larger than the critical threshold frequency $\omega_T$, no matter how large the driving force amplitude is, the complicated solutions cannot exist. As mentioned before, for a fixed damping constant $k$, we obtain a threshold frequency $\omega_T$. Consequently, we find quantitatively that the critical threshold frequency $\omega_T$ at the first branch, $n = 1$, is dependent on the damping constant $k$ with a simple relation:

$$\omega_T = \omega_0 \cdot \left[1 - \left[\frac{k}{k_0}\right]^2\right] \qquad (5)$$

where $\omega_0 = 1.450$ and $k_0 = 1.681$. In the Fig.4a, it shows the relation between the critical threshold frequency, $\omega_T$, and damping constant, $k$ (both the numerical data, •, and the fitting function, —— ). It gives the clear-cut borderline to differentiate the regions of the complicated solutions and the simple ones. According to this relation, it easily reveals that chaos does not occur not only in the overdamped case, $k >> 1$, but also in the underdamped case $\omega >> k$. Although these results have been ever qualitatively suggested by McDonald and Plischke and by Kautz previously, we herein can quantitatively predict it from Eq.(5) that if the damping constant, $k$, excesses a threshold value, $k_0$, then no chaotic behavior can be found. Therefore, we emphasize that the Eq.(5) provides a valuable information of determination of the non-chaotic states in the RSJ model quantitatively.

According to the self-similar feature mentioned before, we further conjecture that each maximum driving threshold frequency of the $n$-th branches, $\omega_{Tn}$, depends on the damping constant k as the critical threshold frequency $\omega_T$ (for the branch $n = 1$) does. By Eq.(5), we guess that it might be scaled as the following:

$$\omega_{Tn} = \omega_T \cdot n^{-\delta} = \omega_0 \cdot \left[1 - \left[\frac{k}{k_0}\right]^2\right] \cdot n^{-\delta} \qquad (6)$$

where $\delta$ is a parameter dependent on the damping constant $k$. In what follows the equation will be verified tentatively in order to make it appear plausible.

Firstly, the respective threshold frequencies of the $n$-th branches, $\omega_{Tn}$'s, are traced out under one fixed damping constant $k$ by the driving amplitude scanning method ($F$ scanning). Then, changing the damping constant $k$ to another one, a new set of the threshold frequencies, $\omega_{Tn}$'s, therefore, can also be found out in the same fashion. After collecting all of the $\omega_{Tn} - n$ relations depending on various values of the damping constant $k$, we summarize them in Fig.4b. In the log-log coordinate, we find the linear relations between the logarithm of the threshold frequency $\omega_{Tn}$ and of the branch number $n$ accordingly. The property of $\omega_{Tn}$ and $n$ accordingly enables us to determine its functional equation :

$$\omega_{Tn} = \omega_c(k) \cdot n^{-\delta} \qquad (7)$$

where the function $\omega_c(k)$ is dependent on the damping constant $k$ and the exponent d is the slope of the straight lines in Fig.4b for the corresponding damping constant $k$. Note that such simple power law relation apparently reveals the nature of renormalization, i.e. it reproduces itself upon rescaling.

In order to verify the existence of Eq.6, we are still necessary to examine the properties of $\omega_c(k)$ in greater details. It immediately raises a question that whether the function $\omega_c(k)$ for arbitrary $n$ can be taken as the functional form $\omega_T = \omega_0 \cdot \left[1 - (k/k_0)^2\right]$ for $n = 1$, in Eq.(6). To this end, we depict the Fig.4c. The content of the figure is clearly shown in a manner that the property of the function $\omega_c$ (as expressed in ●) is consistent with that of the functional form $\omega_T = \omega_0 \cdot \left[1 - (k/k_0)^2\right]$ (as expressed in ——) successfully. Note, herein, that the exactness of Eq.6 is totally a consequence of the fact concerning the emergence of the self-similar feature under the nonlinear dynamics through the RSJ model.

Following the spirit of the physical implication of the renormalization method, we attempt to explore the features of the exponent $\delta$ to the extent at present. Since the value of the exponent $\delta$ depends on the damping constant $k$ (as in Fig.4d), we find the three different cases in the following. For the case of the small damping constant $k$ ($\leq 0.1$), the exponent $\delta$ reaches a constant ($\simeq 2.589$) and is nearly independent of the damping constant $k$. As the case of the damping constant $k$ getting larger and larger ($\geq 0.5$), the increasing of exponent $\delta$ behaves as hyper tangent form and then reaches to its maximum. Finally, the case of right after the maximum, the exponent $\delta$ drops rapidly to zero as the value of the damping constant $k$ approaching to the threshold value, $k_0$. All these results of the exponent $\delta$ can be utilized to realize the self-similar behavior quantitatively.

## B. Duffing Oscillators

The generalized rf-biased Duffing oscillator governed by the equation is given as:

$$\ddot{x} + k\dot{x} + \frac{dV(x)}{dx} = F \sin \omega t \tag{8}$$

where the overdot denotes the derivative with respect to time $t$, $k$ is the damping factor, and $V(x)$ is an anharmonic potential function. This equation has been utilized to model a wide variety of physical systems such as the optical bistability in the multiple-photon absorption process, soft and hard springs, buckled beam, four wave interaction, and plasma oscillation[22–25], et. al.. In general, the potential function $V(x)$ is described by :

$$V(x) = \frac{\alpha}{2}x^2 + \frac{\beta}{3}x^3 + \frac{\gamma}{4}x^4 \tag{9}$$

with $\alpha$, $\beta$, and $\gamma$ coefficients. Actually, by means of some transformations[26], it embraces four fundamental types of potential: $V_1 = \frac{1}{2}x^2 - \frac{1}{3}x^3$, $V_2 = -\frac{1}{2}x^2 + \frac{1}{4}x^4$, $V_3 = \frac{1}{2}x^2 - \frac{\gamma}{4}x^4$, and $V_4 = \frac{1}{2}x^2 + \frac{1}{4}x^4$. The features of the state diagrams of these four types are listed, respectively, as follows;

- For the case of $V_1$ potential ($V_1 = \frac{1}{2}x^2 - \frac{1}{3}x^3$)[27,28]:

The state diagram is shown in Fig.5a. The transition boundaries include hysteresis, period-doubling (PD), crisis, and intermittency. In the state diagram, we also note that the curve of PD folds back at a frequency $\omega_T$. The complicated solutions exist only with the parameters w below the threshold value $\omega_T$. According to our experiments as same as that of the RSJ model, the threshold frequency $\omega_T$ is found to be a function of the damping constant $k$ with a simple form,

$$\omega_T = \omega_0 \left[ 1 + c_1 \left(\frac{k}{k_0}\right) - c_2 \left(\frac{k}{k_0}\right)^2 + c_3 \left(\frac{k}{k_0}\right)^3 - c_4 \left(\frac{k}{k_0}\right)^4 \right] \qquad (10)$$

where $\omega_0 = 2.226$, $k_0 = 2.200$, $c_1 = 0.101$, $c_2 = 4.011$, $c_3 = 4.495$, and $c_4 = 1.585$ (see Fig.5b). $k_0$ is required to satisfy $\omega_T(k_0) = 0$. Moreover, the non-trivial coefficients of odd order of $(k/k_0)$, $c_1$ and $c_3$, are induced by the asymmetrical property of potential well with respect to the origin, while they are equal to zero in the case of symmetrical ones.

- For the case of $V_2$ potential ($V_2 = -\frac{1}{2}x^2 + \frac{1}{4}x^4$)[29,30]:

Fig.6a shows the corresponding state diagram. The potential well $V_2$ has two local wells separated by a bump in the middle of $V_2$. As regards allowing the motion in one of the local wells, the shape of the transition boundaries looks like a swallow tail. The curves of period-doubling are folded. The threshold frequency $\omega_T$ is found to be dependent on the damping constant $k$. And this relation can be fitted to satisfy the following equation:

$$\omega_T(k) = \omega_0 \left[ 1 - \left[\frac{k}{k_0}\right]^2 \right] \qquad (11)$$

where the constants $\omega_0 = 3.218$ and $k_0 = 2.050$ (see Fig.6b). If the driving frequency is beyond the threshold frequency $\omega_T$ and the driving force amplitude does not exceed the boundary $H_{up}$, the complicated solution cannot occur. Eq.11 provides a crucial condition to determine whether the solution is simple or complicated. All these observed features in this system are the same as those of the RSJ model. With further increasing the excitation amplitude up to curve $H_{up}$, the solution becomes stable with the swing throughout two valleys. In this situation the dynamics of the swing closely resembles the case of the infinitely bounded potential $V_4$, while the small effect of the bump in the well is negligible.

- For the case of $V_3$ potential ($V_3 = \frac{1}{2}x^2 - \frac{\gamma}{4}x^4$)[31]:

The state diagram for the transition boundaries with the swallow-tailed form is shown in Fig.7a with damping constant $k = 0.1$. the transitions include hysteresis, symmetry breaking (SB), PD, crisis and intermittency. Due to the fact that the potential barrier is finite, the response will escape from the well after a sequence of bifurcation of Feigenbaum period-doubling and/or intermittency routes to chaos. The threshold frequency, $\omega_T$, at which the curve of PD folds back is well fitted in term of the damping constant $k$ as the following relation:

$$\omega_T(k) = \omega_0 \left[ 1 - \left[ \frac{k}{k_0} \right]^2 \right] \tag{12}$$

where the constants $\omega_0 = 1.528$ and $k_0 = 1.985$; (see Fig.7b). With the driving frequency higher than the threshold frequency, $\omega_T$, only the simple period-1 solutions are obtainable.

- For the case of $V_4$ potential ($V_4 = \frac{1}{2}x^2 + \frac{1}{4}x^4$)[32–35]:

The potential is symmetrical, without an inflection point, and is infinitely bounded as $|x| \to \infty$. The state diagram is shown in Fig.8. The shapes of the transition boundaries are no longer like the swallow tail and can be classified into two groups with characteristic shapes associated with odd and even resonances. The chaotic solutions exist in the even resonant regions (marked as "C" in Fig.8). There is no such threshold frequency $\omega_T$ observed. Also, there is no appearance of the similar $\omega_T - k$ functional relation function in a closed form.

3. **Conclusion**

In this work, we have systematically examined the rf-biased Josephson junction and the Duffing oscillators in a wide range of the parameter space. The crucial formulas of the functional relation, $\omega_T$ vs. $k$, have been derived to give the clear-cut criteria for identification of the existence of the non-chaotic states. Together with our preceding report[26], it is worthwhile mentioning that the functional $\omega_T - k$ relation can be obtained in those potentials which possess the inflection points only. Furthermore, for those potential wells possessed symmetrical property with respect to the origin of the potential, the $\omega_T - k$ relation is in the form of $\omega_T = \omega_0 \left[1 - \left[\frac{k}{k_0}\right]^2\right]$; while asymmetrical, the $\omega_T - k$ relation in the form of $\omega_T = \omega_0 \left[1 + c_1 \left(\frac{k}{k_0}\right) - c_2 \left(\frac{k}{k_0}\right)^2 + c_3 \left(\frac{k}{k_0}\right)^3 - c_4 \left(\frac{k}{k_0}\right)^4\right]$. In addition, the embedded feature of the self-similarity in the RSJ model is also extended to obtain the power law relation between the threshold frequency of each branch, $\omega_{Tn}$, and the branch number, $n$. The threshold frequency of each branch, $\omega_{Tn}$, is rescaled by the branch number $n$ as $\omega_{Tn}(k) = \omega_0 \left[1 - \left[\frac{k}{k_0}\right]^2\right] \cdot n^{-\delta}$. According to the physical implication of the renormalization method conventionally, the exponent $\delta$ has been found to present different kinds of behaviors under three stages of the damping constant $k$, respectively. The generic relation connecting $\omega_T$ and $k$ of the above equations provide a crucial guide post to determine which kinds of solutions (simple or complicated) can be taken in accord with the strength of the frequency at a constant damping in the rf-biased nonlinear oscillators quantitatively.

Finally, we do emphasize that the crucial formulas, $\omega_T = \omega_0 \left[1 - \left[\frac{k}{k_0}\right]^2\right]$ for potentials possessing symmetrical property and $\omega_T = \omega_0 \left[1 + c_1 \left(\frac{k}{k_0}\right) - c_2 \left(\frac{k}{k_0}\right)^2 + c_3 \left(\frac{k}{k_0}\right)^3 - c_4 \left(\frac{k}{k_0}\right)^4\right]$ for asymmetrical potentials, can be tailored the following applications intuitively: (1) we can choose somewhat exact solutions of the system in a chaos-free region with a reference to the restriction of the parameters to obey such crucial formulas. For example, the symmetrical solution in the chaos-free region can be almost precisely expressed in the form of the single mode as $x = a \cdot \sin(\omega t + \phi)$. (2) if we know how to exactly locate where the chaotic region is, then we can plan immediately concerning how to avoid the chaotic noise or to control the chaos in these nonlinear oscillating systems.

## 4. Acknowledgment

This work was supported by the National Science Council of the Republic of China (Taiwan) under contract No. NSC83-0208-M009-37.

**5. Figure Captions:**

Fig.1 The state diagram of RSJ model with $k = 1.0$. The shaded area denotes the complicated behavior regions including those caused by various kinds of bifurcation. There is no chaotic behaviors found as the rf frequency exceeds $\omega_T$. The inset is the corresponding potential well, $1 - \cos x$. The digits from 1 to 7 represent the number of the local wells which the motion runs. The phase portraits of the states marked from 'a' to 'g' are presented in the figure 3. The symbols of "S" and "SB" represent the symmetrical and asymmetrical solutions, respectively.

Fig.2 There bifurcation diagrams of RSJ model with $k = 1.0$ and F from 0.0 to 8.0 and for $\omega = 0.313, 0.400$, and $0.600$, respectively. All the responses on the Poincaré section are carried into the intervals of $[-\pi, \pi]$ by the transform of $x \pm 2m\pi$. The similar scenarios of bifurcation appear itself again and again in each branch.

Fig.3 The phase portraits with $k = 1.0$, $\omega = 0.313$, and (a)$F = 1.00, n = 1$; (b)$F = 1.75, n = 2$; (c)$F = 2.60, n = 3$; (d)$F = 3.55, n = 4$; (e)$F = 4.50, n = 5$; (f)$F = 5.60, n = 6$; (g)$F = 6.50, n = 7$; (h)$\omega = 0.05, F = 8.00, n = 51$. The value of $n$ represents the number of the local wells which the motion runs.

Fig.4a The numerical simulation data of $\omega_T - k$ compared with the fitting function, $\omega_T = \omega_0[1 - (k/k_0)^2]$, where $\omega_0 = 1.450$ and $k_0 = 1.681$.

Fig.4b The numerical simulation data $\omega_{Tn} - n$ for various values of $k$ and fitted by the functional relation $\omega_{Tn} = \omega_c n^{-\delta}$.

Fig.4c The dependence on $k$ of fitting constant $\omega_c$ obtained from fig.4(b).

Fig.4d The dependence on $k$ of fitting constant $\delta$ obtained from fig.4(b).

Fig.5 (a) The state diagram of asymmetrical one-well Duffing oscillator including the primary ($A_I$) and subharmonic ($B_I$) resonance regions. $C$ denotes the complicated solution region enclosed by period-doubling and reverse period-doubling transition boundaries and $E_s$ escaping boundary. $\omega_T$ indicated the maximum rf frequency for chaotic behaviors existing. The potential well is drawn in the inset. (b) The $\omega_T - k$ relation, $\omega_T(k) = \omega_0[1 + c_1(k/k_0) - c_2(k/k_0)^2 + c_3(k/k_0)^3 - c_4(k/k_0)^4]$ with $\omega_0 = 2.226, k_0 = 2.200, c_1 = 0.101, c_2 = 4.011, c_3 = 4.495$, and $c_4 = 1.585$. The black circles denote the experimental simulation data and the solid line for fitting.

Fig.6 (a) The state diagram of symmetrical two-well Duffing oscillator including the primary ($A_I$) and subharmonic ($B_I$) resonance regions. $C$ denotes the complicated solution region enclosed by period-doubling and reverse period-doubling transition boundaries and H$_{up}$ the jump-up boundary. The potential well is presented in the inset. As long as the rf frequency is above $\omega_T$ and the rf amplitude is below H$_{up}$ transition boundary, there is no chaotic behaviors found. (b) The $\omega_T - k$ relation (9), $\omega_T = \omega_0[1 - (k/k_0)^2]$ with $\omega_0 = 3.218$ and $k_0 = 2.050$, of the rf-biased Duffing oscillator with double well potential.

Fig.7 (a) The state diagram of symmetrical one-well Duffing oscillator including the primary ($A_I$) and subharmonic ($B_I$) resonance regions. $C$ denotes the complicated solution region enclosed by period-doubling and reverse period-doubling transition boundaries and $E_s$ escaping boundary. The potential well is presented in the inset. As long as the rf frequency is above $\omega_T$, there is no chaotic behaviors found. (b) The $\omega_T - k$ relation (10), $\omega_T = \omega_0[1 - (k/k_0)^2]$ with $\omega_0 = 1.528$ and $k_0 = 1.985$, of the rf-biased Duffing oscillator with one well potential.

Fig.8 The state diagram of the Duffing oscillator with infinite potential well. The presented transition boundaries are including saddle-node ($H$), symmetry-breaking ($AS$), and period-doubling and reverse period-doubling ($C$) bifurcation. The shaded terrain de-notes the complicated solution regions. There is no $\omega_T$ found.